\documentclass[preprint,showpacs,preprintnumbers,amsmath,amssymb]{revtex4}

\usepackage{epsfig}

\def\eg{{\it e.g.}\ } 
\def\etal{{\it et al.}\ } 
\def\ie{{\it i.e.}\ }
 
\def\bB0{{\bf B}_0}
\def\kpe{k_{\perp}}
\def\pa{\parallel}
\def\kpa{k_{\parallel}}
\def\be{\begin{equation}}
\def\ee{\end{equation}}

\usepackage{color}
\def\add#1{{#1}}

\begin{document}
\preprint{1}

\title{On \add{spectral} scaling laws for incompressible anisotropic magnetohydrodynamic 
turbulence}

\author{S\'ebastien Galtier}

\affiliation{Institut d'Astrophysique Spatiale, Universit\'e 
de Paris-Sud, CNRS, B\^atiment~121, 91405 Orsay Cedex, France}

\author{Annick Pouquet}
\affiliation{IMAGe (CISL/NCAR), P.O. Box 3000, Boulder, CO 80307-3000, USA}

\author{Andr\'e Mangeney}
\affiliation{LESIA, B\^atiment 16, Observatoire de Paris-Meudon, 
5, Place Jules Janssen, 92195 Meudon Cedex, France}

\date{\today}

\begin{abstract}
A heuristic model is given for anisotropic magnetohydrodynamics (MHD) turbulence in 
the presence of a uniform external magnetic field $B_0 \, {\bf {\hat e}_{\pa}}$. 
The model is valid for both moderate and strong $B_0$ and is able to describe both the 
strong and weak wave turbulence regimes as well as the transition between them. The main 
ingredient of the model is the assumption of constant ratio at all scales between 
\add{the} linear wave period and \add{the} nonlinear turnover timescale. Contrary to 
the model of critical balance introduced by Goldreich and Sridhar [P. Goldreich and S. 
Sridhar, ApJ {\bf 438}, 763 (1995)], it is not assumed in addition that this ratio be 
equal to unity at all scales which allows us to use the Iroshnikov-Kraichnan 
phenomenology. It is then possible to recover the widely observed anisotropic scaling 
law $\kpa \propto \kpe^{2/3}$ between parallel and perpendicular wavenumbers (with 
reference to $B_0 \, {\bf {\hat e}_{\pa}}$) and to obtain the universal prediction, 
$3\alpha + 2\beta = 7$, for the total energy spectrum 
$E(\kpe,\kpa) \sim \kpe^{-\alpha} \, \kpa^{-\beta}$. In particular, with such a 
prediction the weak Alfv\'en wave turbulence constant-flux solution is recovered and, 
for the first time, a possible explanation to its precursor found numerically by 
Galtier \etal [S. Galtier et al., J. Plasma Phys. {\bf 63}, 447 (2000)] is given. 
\end{abstract}


\maketitle

\section{Introduction}

Turbulent flows are often studied under the assumptions of homogeneity and isotropy 
(see, for a review, Ref. \cite{frish,lesieur}). Such assumptions are convenient 
for theoretical studies but are not always justified physically. For example, it 
is well known that stratification or rotation applied to neutral flows lead 
to anisotropic turbulence (see \eg Ref. \cite{godeferd,galt03i}). Isotropy is 
even more difficult to justify in astrophysics where a magnetic field is almost 
always present at the largest scale of the system. The magnetohydrodynamics 
(MHD) approximation has proved to be quite successful in the study of a variety 
of space plasmas. During the last quarter of century many studies have been 
devoted to incompressible MHD turbulence in the presence of a uniform external 
magnetic field $B_0 \, {\bf {\hat e}_{\pa}}$ (see Ref. \cite{leh55,strauss76,gaps81,
mont81,shebalin,oughton94,goldreich95,ng96,kinney,matthaeus98,cho2000,galt00,milano2001,
maron2001,galt02}). 
One of the most clearly established results is that the presence of $B_0$ leads 
to a bi-dimensionalization of an initial isotropic energy spectrum: the turbulent 
cascade transfers energy preferentially to perpendicular wavenumbers, \ie in 
the direction transverse to $B_0 \, {\bf {\hat e}_{\pa}}$. 

\medskip

Constant-flux spectra are known to occur in many instances in turbulent flows, 
the best example of which being the Kolmogorov energy spectrum following a 
$E(k) \propto k^{-5/3}$ law for three-dimensional Navier-Stokes turbulence (see \eg 
Ref. \cite{K41a}). Power law spectra are also measured in turbulent MHD flows but 
the value of the scaling index is still hardly discussed in the community. 
The first prediction in MHD was given independently by Iroshnikov and Kraichnan in 
Ref. \cite{iro,kraichnan65} (hereafter IK). They argued that the destruction of 
phase coherence by Alfv\'en waves traveling in opposite directions along the 
local large scale magnetic field introduces a new timescale and a slowing down 
of energy transfer to small scales. Assuming isotropy, the dimensional analysis 
for three wave interactions leads to a $E(k) \propto k^{-3/2}$ spectrum for 
the total energy (see also Ref. \cite{matthaeus89}). Many direct numerical 
simulations of strong turbulence in isotropic ($B_0=0$) MHD have been made 
during the last years (see \eg Ref. \cite{politano89,cho2000,maron2001}) but a 
definitive conclusion about the spectrum index is still not achieved mainly 
because the Kolmogorov and IK predictions are very close; furthermore, such 
scaling laws may be slightly altered by intermittency effects, and the 
numerical resolution is barely sufficient to determine such spectral indices. 
Goldreich and Shridar in Ref. \cite{goldreich95} proposed in 1995 a heuristic model 
of strong turbulence for anisotropic (moderate $B_0$) MHD where the distinction 
between the perpendicular ($k_{\perp}$) and parallel ($k_{\parallel}$)
wavenumbers is made. This model is based on a critical balance between linear 
wave periods $\tau_A$ and nonlinear turnover timescales $\tau_{NL}$ for which 
the equality $\tau_A = \tau_{NL}$ is assumed to hold at all scales in the 
inertial range. (Only the symmetric case, for which the Alfv\'en waves 
traveling in opposite directions carry equal energy fluxes, is considered here 
and in the remainder of the paper.) Then the one dimensional perpendicular 
power spectrum for the total energy scales as $E(\kpe) \propto \kpe^{-5/3}$ 
whereas the parallel and perpendicular spatial sizes of eddies are correlated 
according to the scaling law $\kpa \propto \kpe^{2/3}$. The latter prediction 
is rather new (see however Ref. \cite{higdon}) and seems well observed in recent 
direct numerical simulations for moderate $B_0$ (see Ref. \cite{cho2000,maron2001}). 

\medskip

As is well known, in the limit of large $B_0$ MHD turbulence becomes strongly 
anisotropic \add{and mainly weak in the sense that, in Fourier space, the domain 
of applicability of strong turbulence is confined to a region localized around 
the $\kpa=0$ plane.} The formalism of weak Alfv\'en wave turbulence developed by 
Galtier et al. in Ref. \cite{galt00,galt02} is well adapted to this situation. 
It leads to the so--called wave kinetic equations for which the exact power law 
solution for the total energy is $E(\kpe, \kpa) \propto \kpe^{-2} f(\kpa)$. 
The function $f(\kpa)$ is undetermined because of the dynamical decoupling of 
parallel planes in Fourier space (this is the signature of the absence of 
energy transfer along the $B_0$ direction; it thus represents a shadow of the 
initial conditions). Numerical simulations of the wave kinetic equations show 
clearly such a constant-flux spectrum but it also reveals the existence of a 
transient spectrum during the front propagation towards small scales with a 
steeper power--law in $k_{\perp}^{-7/3}$, the dynamics of which is not yet 
clarified (see \eg Ref. \cite{falko91,newell01}). The discovery of such transient 
spectra in wave turbulence and the possible existence of a family of solutions 
that are not caught by the usual technique of conformal transform (given \eg in 
Ref. \cite{ZLF}) constitute a new exciting topic of research where some progress 
is currently being made (see, for example, Ref. \cite{colm04,lvov04}). When using 
a shell model of strong turbulence, it has also been found in Ref. \cite{melander} 
that, when considering the decay of energy 
in time as a power law, initial transients occur that also follow power--laws 
to leading order and that precede the final power--law decay of the energy; 
the origin of such transients in time may well be a transient in the Fourier 
energy spectrum preceding the establishment of a Kolmogorov--like spectrum, 
although this point has not been documented yet. 

\medskip

In this paper, we propose a heuristic model that describes anisotropic MHD flows for 
the regimes of strong (moderate $B_0$) and weak wave (strong $B_0$) turbulence as well 
as the transition between them. As a result of our analysis, a family of solutions 
is found for the anisotropic total energy spectrum from which the transient spectrum 
described above is a particular solution. We show that the model supports the same 
anisotropic scaling law between the parallel and perpendicular wavenumbers as the 
one found in the context of critical balance but it is more general in the sense that 
here we do not impose equality between linear wave periods and nonlinear turnover 
timescales which allows us to use the IK phenomenology. We finally propose to extend 
the anisotropic model to two other types of fluids.

\section{Anisotropic MHD model}

The presence of the mean 
magnetic field $B_0$ leads to anisotropy and thus to different variations with 
wavenumber directions, leading us to distinguish between ${\bf \kpe}$ and 
${\bf \kpa}$. More precisely we will assume that $\kpe \gg \kpa$, {\it i.e.} 
that under the external agent (here, ${\bf B}_0$), the turbulent flow develops 
principally in the direction perpendicular to this agent. We will consider the 
symmetric case for which, in particular, the {\it r.m.s.} value \add{at any} 
scale in the inertial range of the 
fluctuating velocity field ${\bf v}$ and magnetic field ${\bf b}$ have the same 
order of magnitude; note that ${\bf b}$ is taken in velocity units since we 
are restricting the analysis to the incompressible MHD case (the density is 
constant and can be assumed to be equal to unity). 

\medskip

\begin{figure}[h]
\centerline{\hbox{\psfig{figure=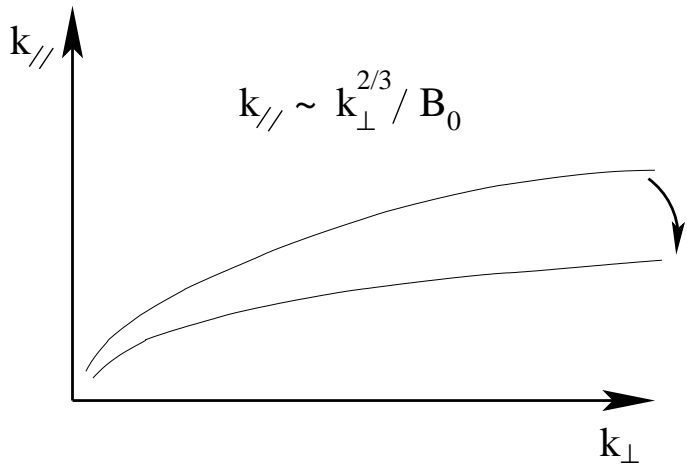,width=9cm}}}
\caption{Path, in Fourier space, followed by the MHD turbulence dynamics for two 
different values of the external magnetic field $B_0$. A higher value of $B_0$ 
leads to weakening of parallel transfers with a curve closer to the $\kpe$-axis. 
When $B_0$ is infinite the path identifies with the $\kpe$-axis and no transfers 
are possible along the parallel direction. Note that the curves are not extended 
down to the origin since the anisotropic law is only valid in the inertial range.}
\label{path}
\end{figure}

In the classical Kolmogorov  phenomenology (hereafter, K41), the fluctuations are 
distributed isotropically and there is only one timescale, the nonlinear time or 
eddy turnover time $\tau_{NL}$, which is also the transfer time of the energy to 
small scales within the system, \ie $\tau_{tr}=\tau_{NL}$. The rate of energy 
transfer per unit mass writes ${\cal E}_{K41} \sim E / \tau_{tr}$, where $E$ is 
the total \add{(kinetic plus magnetic)} energy \add{at a given scale}. In the very 
same spirit, we can develop the IK phenomenology for 
anisotropic MHD turbulence using explicitly the fact that the eddy turnover time 
is $\tau_{NL}\sim (v k)^{-1} \sim (v k_{\perp})^{-1}$ and the Alfv\'en wave period 
is $\tau_A \sim 1/(\kpa B_0)$. We assume that these two timescales are {\it not} 
equal which allows us to use the IK phenomenology. The rate of energy transfer 
per unit mass now writes 
\be
{\cal E}_{IK_a} \sim \frac{E}{\tau_{tr}} \sim \frac{v^2}{\tau_{tr}} \, ,
\label{relation1}
\ee
where the transfer time will be assumed to scale as 
\be
\tau_{tr} = \tau_{NL} \, \frac{\tau_{NL}}{\tau_A} \, , 
\label{relation2}
\ee
as is known to be the case for three-wave interaction processes (see \eg 
Ref. \cite{iro,kraichnan65,galt00}). The 
subscript ``a'' in $IK_a$ stands for the anisotropic version of IK. 
Including relation (\ref{relation2}) into (\ref{relation1}), one obtains
\be
{\cal E}_{IK_a} \sim {v^4 \kpe^2 \over B_0 \, \kpa} \, .
\label{epsIK} 
\ee
We now define the anisotropic energy spectrum, 
\add{assuming self-similarity in both ($\perp,\ \parallel$) directions:}
\be
E(\kpe,\kpa) \sim \kpe^{-\alpha} \, \kpa^{-\beta} \, ,
\label{alphabet}
\ee
where $\alpha$ and $\beta$ are unknown. It is a 2D energy spectrum. In other words, 
the total energy \add{of the system} is recovered by directly integrating the 
spectrum along $\kpe$ and $\kpa$, \ie 
$E^{\add{sys}} = \int \int E(\kpe,\kpa) \, d\kpe d\kpa$ (see \eg Ref. \cite{galt00}, 
for a rigorous definition). It is important to note at this stage a difference with 
the classical phenomenology where the energy spectrum is not introduced with unknown 
indices but rather deduced by the analysis. \add{The phenomenology proposed to describe 
MHD turbulence is, of course, not unique; one can find recent approaches 
in \eg Ref. \cite{boldyrev05,zhou05}. As is well known, the presence of an 
external magnetic field ${\bf B}_0$ leads to a reduction of nonlinear transfers along 
its direction which, in turns, leads to a reduction in size in Fourier space
of the inertial range. Numerical simulations, for the most part at moderate resolutions, 
and/or using the Reduced MHD approximation (see {\it e.g.} Ref. \cite{Strauss79}) show 
indeed that a strong $B_0$ leads to an absence of inertial range in $\kpa$ with a 
spectrum mainly exponential (see \eg Ref. \cite{kinney,Mont87,cho2002,oughton04}) as it 
is usually observed in turbulence at large wavenumbers. 
Since heuristic models deal with power laws, they are mainly able to describe what 
happens inside the inertial range, without saying much about the strength of the 
nonlinear transfers and thus about the size of the inertial range. 
We are assuming, as noted before (see equation (4)) that power-law develops in 
$\kpa$ as well, {\it i.e.} that the Reynolds number is extremely large, as found in 
astrophysical flows; in particular, it may need to be substantially higher than in the 
fully isotropic case since transfer in the $k_{\parallel}$ direction is strongly impeded 
in the presence of a uniform magnetic field.
Therefore, relation (\ref{alphabet}) introduced above gives a description of the possible 
inertial ranges in both $\kpe$ and $\kpa$ directions; additionally, it offers an 
opportunity to describe continuously the phase transition between the strong and weak 
turbulence regimes.} 
We introduce relation (\ref{alphabet}) into (\ref{epsIK}), and by noticing that 
$v^2 \sim E \sim E(\kpe,\kpa) \, \kpe \kpa$, we obtain:
\be
B_0 \, {\cal E}_{IK_a} \sim \kpe^{4-2\alpha} \, \kpa^{1-2\beta} \, ,
\ee
which can also be rewritten in terms of a ($\kpe$, $\kpa$) relationship as:
\be
\kpa \sim 
(B_0 \, {\cal E}_{IK_a})^{1 \over 1-2\beta} \, \kpe^{4-2\alpha \over 2\beta-1} \, .
\label{ani2} \ee
This is our first anisotropic relation. 
In order to proceed further, we are now seeking a second relationship between the two 
scaling indices $\alpha$ and $\beta$. One way to obtain a unique relation between 
them is to use the assumption of constant ratio between $\tau_{NL}$ and $\tau_A$ at 
all scales \add{in the inertial range}. In other words, we will assume that 
\be
\chi = \frac{\tau_A}{\tau_{NL}}
\label{chi}
\ee
is a constant at all scales but not necessarily equal to one as it is in the critical 
balance model in Ref. \cite{goldreich95}. \add{Condition (\ref{chi}) can be understood 
as a formal balance between the linear and nonlinear terms in the ideal MHD equations.} 
It is important to note that the precise value 
of $\chi$ is not important in itself since it is just a numerical factor that does not 
change the power law scaling. However if $\chi$ is smaller than one we are allowed 
to use the IK phenomenology instead of the K41 one for $\chi=1$. A ratio of one seems 
to be very restrictive and does not correspond to some of the results stemming from 
direct numerical simulations where $\chi$ can be smaller than unity, as observed 
for example in Ref. \cite{ng2003}, a two-dimensional geometrical case where local 
anisotropy is possible, or in solar wind {\it in situ} measurements (see Ref. 
\cite{matthaeus95}) where $\chi$ seems to be smaller than unity. It is also in 
apparent contradiction with wave turbulence theory for which we have $\chi \ll 1$. 
We will see that the assumption of constant ratio at all scales \add{in the inertial 
range} is sufficient to achieve a unified model that describes both strong and weak 
MHD turbulence. With the definition of the timescales given above, we obtain
\be
\chi \sim \frac{v \, \kpe}{B_0 \, \kpa} \, .
\label{relationchi}
\ee
Including expression (\ref{relationchi}) into (\ref{epsIK}) gives: 
\be
\kpa \sim {{\cal E}_{IK_a}^{1/3} \over \chi^{4/3} B_0} \, \kpe^{2/3} \, . 
\label{ani4}
\ee
If we use explicitly the fact that both $\chi$ and the rates of energy transfer per 
unit mass do not depend on the scale (see \eg Ref. \cite{K41a,iro,kraichnan65}), we 
finally obtain
\be
\kpa \sim {\kpe^{2/3} \over B_0} \, . 
\label{ani5}
\ee
This leads us to the first main conclusion of this paper: the same scaling in 
wavenumbers as the critical balance model is obtained when 
$\chi$ is {\it not} equal to unity, \ie when the IK phenomenology is used. 
It thus seems to be a general law for incompressible MHD turbulence, when 
either the Kolmogorov phenomenology for energy transfer or the $IK_a$ three-wave 
formulation of energy transfer is utilized, as long as a critical balance 
(generalized to any ratio differing from unity) is assumed between stretching by 
velocity gradients and wave motions in the presence of a uniform field $B_0$. 
\add{Condition (\ref{ani5}) may be seen as a path in $\kpe$--$\kpa$ Fourier space 
followed naturally by the dynamics for different values of the (imposed) uniform 
magnetic field. An illustration is given in Figure (\ref{path}) for two different 
values of $B_0$. It is interesting to note that condition (\ref{ani5}) is not 
incompatible with the weak wave turbulence prediction since in the infinite 
limit of $B_0$ the curve identifies with the $\kpe$-axis and therefore no transfers 
are possible along the parallel direction.}

\medskip

Another important and somewhat unexpected consequence of this analysis has to 
deal with the scaling of the energy spectrum. In the context of critical balance, 
it is claimed that the energy spectrum derived with an anisotropic Kolmogorov 
phenomenology scales like $E(\kpe) \sim \kpe^{-5/3}$. Here we show that it is in 
fact possible to find a multitude of spectral indices $\alpha$ and $\beta$ that 
satisfy the assumption of constant ratio $\chi$ at all scales in the inertial range. 
Such indices follow in fact a general linear relationship. To find this relation
we need to equalize the power-law behaviors found in relations (\ref{ani2}) and 
(\ref{ani5}) between the parallel and perpendicular distribution of modal energy 
in Fourier space. If we use explicitly the fact that the rates of energy transfer 
per unit mass do not depend on the scale (see \eg Ref. \cite{K41a,iro,kraichnan65}), 
we finally obtain
\be
3\alpha + 2\beta = 7 \, .
\label{ab}
\ee

\medskip

Relationship (\ref{ab}) is general and can be used for strong turbulence as well 
as for weak wave turbulence. 
Note that the relationship (\ref{ab}) is compatible with the anisotropic Kolmogorov 
spectrum (with $\alpha=5/3$ and $\beta=1$) as advocated by Goldreich and Sridhar in 
Ref. \cite{goldreich95}, and with the anisotropic IK-like spectrum corresponding to 
three-wave interactions for weak MHD turbulence ($\alpha=2$ and $\beta=1/2$) (see Ref. 
\cite{goldreich97,ng97,galt00}). We conjecture that it is compatible with the physics 
of the transitional regime between weak and strong MHD turbulence as well. 
In other words the law (\ref{ab}) we just derived shows that the $2/3$ scaling 
between a dependence of $\kpe$ and $\kpa$ is not a unique signature of the Kolmogorov 
spectrum but rather a signature of the rate at which energy is transfered to small 
scales where it is dissipated, \ie a trace of the decay of energy, and as such an 
unavoidable dimensional law \add{assuming equation (\ref{alphabet}) holds and 
that $\chi$ is independent of wavenumbers in the inertial range.}

\section{Discussion}

The preceding remark may be linked to the heretofore unexplained following 
finding: using direct numerical simulations in two space dimensions, it was 
shown in Ref. \cite{gomez99} that the structure functions of order $p$ based on 
the energy flux to small scales (as expressed in terms of the exact laws for 
MHD turbulence derived in Ref. \cite{poli98a,poli98c}) have a self--similar scaling 
in the inertial range which is compatible with the scaling for the velocity 
structure functions in fluid turbulence, whereas the structure functions of 
the basic fields (velocity and magnetic fields) are more intermittent insofar 
as they depart more significantly from a linear scaling with $p$. What this 
paper shows is that both models (K41 and IK) can be seen in a unifying way, 
illustrated by the law (\ref{ab}). 
Note that in the case of the advection of a passive tracer such 
as temperature, it also seems that the scaling of the structure functions 
based on the flux of the tracer is close to the fluid scaling laws (see Ref. 
\cite{boratav,pinton98}), whereas the tracer itself is well--known to be
strongly intermittent.

\medskip

More surprisingly perhaps, the choice of $\alpha=7/3$ and $\beta=0$ also 
satisfies the law (\ref{ab}) derived in this paper. This $\kpe^{-7/3}$ spectrum 
was found in Ref. \cite{galt00} as a precursor to the constant flux solution of the 
Alfv\'en wave kinetic equations which establishes itself later in time; indeed, 
in that paper, the $7/3$ spectrum is found numerically for the case $\kpa=0$, \ie 
without any dependence in $\kpa$ (corresponding to $\beta=0$). 
The simple model proposed here thus sheds some light, albeit heuristically, on
two intriguing facts that have emerged recently concerning weak wave turbulence: 
(i) the fact that, preceding in time the constant flux solution, a power-law 
spectrum (called the precursor) establishes itself the origin of which was unrelated 
to anything known about turbulence spectra (see Ref. \cite{galt00,colm04}) until
now, and (ii) the fact that in some cases (see \eg Ref. \cite{lvov04}) in the weak 
wave turbulence regime, a wealth of power-law solutions in the $(\alpha, \beta)$
plane can be found numerically as stationary solutions to the wave kinetic 
equations, although it is not clear whether such solutions are attractive,
nor whether they are stable. The link between the dimensional argument given 
here which allows to recover the $\alpha=7/3$ precursor spectrum, and the 
self-similar argument given in Ref. \cite{galt00}, which is compatible with the 
$\alpha=7/3$ spectrum, remains to be clarified.  Indeed, in Ref. \cite{galt00} 
one recovers the numerically observed law of self-similar decay of the energy 
spectrum, assuming that the energy spectrum scales in the precursor phase 
as $\kpe^{-7/3}$, whereas in this paper we give a heuristic justification 
to that same $-7/3$ law in terms of dimensional arguments compatible with 
three-wave interactions and with the assumption that there is no $\kpa$ 
dependency in the precursor solution. We show furthermore that this law 
stems from the same type of unified approach that also gives the other known 
spectra for MHD turbulence. In order to ascertain the validity of the model 
derived here, numerical computations with the full three--dimensional MHD 
equations can be envisaged but being able to distinguish between different 
power laws which are close in their respective spectral indices may prove 
difficult. This point is currently investigated and will be presented elsewhere. 

\medskip

The argumentation delineated here could be used for other types of wave 
turbulence, like in the case of whistler waves (see Ref. \cite{galt03}), 
inertial waves (see Ref. \cite{galt03i}) or gravity waves (see Ref. \cite{caillol00}). 
For whistler waves that can be encountered in Hall MHD or electron MHD, the 
characteristic wave period scales as $\tau_W \sim [\kpe \kpa]^{-1}$ (see \eg 
Ref. \cite{dastgeer00}). One finds, following the same analysis as presented 
before in this paper, that $\kpa \sim \kpe^{1/3}$ 
(see also Ref. \cite{lazarian04}); hence, the prediction for scaling exponents 
as defined in equation (\ref{alphabet}) for the whistler turbulence 
energy spectrum becomes: $3\alpha + \beta = 8$.
(Note that here, the eddy turnover time is based on the magnetic field, \ie 
$\tau_{NL} = [v \, \kpe]^{-1} = [b \, \kpe^2]^{-1}$, 
since ${\bf v} \propto \nabla \times {\bf b}$.) 
The known constant--flux solution, $\alpha=5/2$ and $\beta=1/2$, to the 
wave kinetic equations derived in Ref. \cite{galt03} is recovered again as
well as the strong turbulence prediction, $\alpha=7/3$ and $\beta=1$ 
(see \eg Ref. \cite{biskamp}). 

\medskip

When considering inertial waves in rotating turbulence for a Navier-Stokes fluid,
the wave period scales as 
$\tau_I \sim \kpe / \kpa$. The same analysis as before leads to the 
following relationship between scaling exponents: $3\alpha + 5\beta = 10$. 
The known constant--flux solution to the wave kinetic equations written in 
Ref. \cite{galt03i} for wave turbulence corresponds to $\alpha=5/2$ and 
$\beta=1/2$, which does fulfill the above relationship and we recover the 
Kolmogorov prediction for strong turbulence as well, {\it viz.} $\alpha=5/3$ 
and $\beta=1$ for a weak rotation rate (see the comment in Ref. \cite{note4}). 
This prediction again could be tested using direct numerical simulations, 
and the possible existence of precursors could also be studied and checked 
against the compatibility with the above relationship between the scaling 
exponents $\alpha$ and $\beta$ in the rotating case.

\begin{acknowledgments}
We wish to thank Pablo Mininni and H\'el\`ene Politano for interesting 
discussions.
Financial support from the Research Training Network TOSTISP through EEC 
grant HPRN-CT-2001-00310, from PNST/INSU/CNRS, from NSF-CMG grant 0327888, 
and from NASA grant in Geospace Science NRA 03-OSS-01, are all gratefully 
acknowledged.
\end{acknowledgments}


\begin{thebibliography}{}
\bibitem{note4}
Writing in general that the characteristic wave period scales as
$\tau_w \sim k_{\perp}^{\gamma} \  k_{\parallel}^{\delta}$, one finds
$k_{\parallel}\sim k_{\perp}^{-(\gamma-2\alpha+4) / (\delta-2\beta+2)}
\sim k_{\perp}^{-(3 \gamma+2) / 3\delta}$; hence, the generalization of 
equation (\ref{ab}) for waves with the above dispersion relation with the 
turnover time $\tau_{NL}=\ell/u_{\ell}$ is found to be: 
$(3\gamma+2)(1-\beta) = \delta(5-3\alpha)$.
\end{thebibliography}
\end{document}